\documentclass[conference,letterpaper]{IEEEtran}
\IEEEoverridecommandlockouts

\usepackage[utf8]{inputenc} 
\usepackage[T1]{fontenc}
\usepackage{url}
\usepackage{ifthen}
\usepackage{cite}
\usepackage[cmex10]{amsmath} 
\usepackage{amssymb,amsthm}
\usepackage{tikz}
\usepackage{stmaryrd}
\usepackage{acronym}
\usepackage{pgfplots}
\pgfplotsset{compat=1.17}
\usepackage{xcolor}
\usepackage{subfigure}
\usepackage{graphicx} 
 \usetikzlibrary{calc}

\newcommand{\brackets}[1]{\left(#1\right)}
\newcommand{\sbrackets}[1]{\left[#1\right]}

\newtheorem{theorem}{Theorem}

\newtheorem{lemma}[theorem]{Lemma}

\newtheorem{definition}[theorem]{Definition}
\newtheorem{remark}{Remark}

\newcommand{\indep}{\perp \!\!\! \perp}

\acrodef{dmc}[DMC]{discrete memoryless channel}
\acrodef{mmse}[MMSE]{minimum mean square error}
\acrodef{zec}[ZEC]{zero estimation cost}

\begin{document}
\title{Zero Estimation Cost Strategy for Witsenhausen Counterexample with Causal Encoder} 


\author{%
  \IEEEauthorblockN{Mengyuan Zhao and Tobias J. Oechtering}
  \IEEEauthorblockA{Division of Information Science and Engineering \\
                   KTH Royal Institute of Technology\\
                   10044 Stockholm, Sweden
                   }
  \and
  \IEEEauthorblockN{Maël Le Treust}
  \IEEEauthorblockA{Univ. Rennes, CNRS, Inria, IRISA UMR 6074\\
                    35000 Rennes, France
                    }
\thanks{This work is supported by Swedish Research Council (VR) under grant 2020-03884. The work of Maël Le Treust is supported in part by PEPR NF FOUNDS ANR-22-PEFT-0010.} 
}

\maketitle


\begin{abstract}
We propose a \ac{zec} scheme for causal-encoding noncausal-decoding vector-valued Witsenhausen counterexample based on the coordination coding result.
In contrast to source coding, our goal is to communicate a controlled system state.  The introduced ZEC scheme is a joint control-communication approach that transforms the system state into a sequence that can be efficiently communicated using block coding. The noncausal decoder receives sufficient information for reconstructing the system state perfectly, enabling the achievable estimation cost to be zero. Numerical results show that our approach significantly reduces the power budget required for achieving zero-estimation-cost state reconstruction at the decoder. In the second part, we introduce a more general non-zero estimation cost (Non-ZEC) scheme. We observe numerically that the Non-ZEC scheme operates as a time-sharing mechanism between Witsenhausen's original two-point strategy and the ZEC scheme. Overall, by leveraging block-coding gain, our proposed methods substantially improve the power-estimation trade-off for Witsenhausen counterexample.
\end{abstract}

\section{Introduction}

In 1968, Witsenhausen proposed his renowned counterexample, highlighting the suboptimality of affine strategies in the Linear Quadratic Gaussian (LQG) settings with non-classical information pattern \cite{witsenhausen1968}. This counterexample has since become a prominent toy example in the study of distributed decision-making \cite{bansal1986stochastic, silva2010control, yuksel2013stochastic, gupta2015existence} and information-theoretic control \cite{martins2005fundamental, freudenberg2008feedback, derpich2012improved, AgrawalDLL15, Akyol2017information, charalambous2017hierarchical, wiese2018secure, Stavrou2022sequential}.

The vector-valued extension of Witsenhausen counterexample \cite{Grover2010Witsenhausen} facilitates the application of many information-theoretic approaches \cite{elgamal2011nit, kim2008state,sumszyk2009information, choudhuri2013causal} to analyze this open problem. Among these, the coordination coding method \cite{cuff2010coordination, cuff2011hybrid, cuff2011coordination} has proven powerful in building cooperative behavior among different decision-makers, yielding novel bounds and insights into distributed decision-making problems.

Recent advances in this direction for Witsenhausen counterexample are the single-letter characterizations for the optimal power-estimation cost trade-off region in various causal decision-making frameworks, including causal decoding \cite{Treust2024power}, causal coding with feedback \cite{zhao2024causal}, and, in particular, causal encoding \cite{zhao2024coordination}. The auxiliary random variables (aux. RVs) involved in the single-letter expressions, not only capture the asymptotic behavior of the costs, but also explicitly carry the dual role of control in Witsenhausen counterexample: joint state control and information communication.

Building upon this idea, the design of aux. RVs that can be efficiently communicated becomes a key question. In \cite{zhao2024CDC}, we explored control designs within the class of jointly Gaussian aux. RVs and determined the optimal estimation cost as a function of power cost. This optimal Gaussian scheme is shown to operate as a time-sharing mechanism between two affine strategies. However, it is outperformed by Witsenhausen's two-point strategy, which communicates the controlled system state more efficiently by designing it to be a binary sign symbol. 

\begin{figure}[t]
  \centering


\begin{tikzpicture}[scale=0.86]
    \draw (2,0) rectangle (3,1);
    \draw (6.8,0) rectangle (7.8,1);

    \draw (4.2,0.5) circle (0.2) node {$+$};
    \draw (5.6,0.5) circle (0.2) node {$+$};

    \filldraw (1,1.5) circle (2pt) node[above] {$X_{0,t}\sim \mathcal{N}(0,Q)$};
    \filldraw (5.6,1.5) circle (2pt) node[above] {$Z_{1,t}\sim \mathcal{N}(0,N)$};

    \draw[->] (1,1.5) -- (1,0.5) -- (2,0.5);
    \draw[->] (1,1.5) -- (4.2,1.5) -- (4.2,0.7);
    \draw[->] (3,0.5) -- (4,0.5);
    \draw[->] (4.4,0.5) -- (5.4,0.5);
    \draw[->] (4.9,0.5) -- (4.9,-0.5) -- (8.8,-0.5);
    \draw[->] (5.6,1.5) -- (5.6,0.7);
    \draw[->] (5.8,0.5) -- (6.8,0.5);
    \draw[->] (7.8,0.5) -- (8.8,0.5);

    \node at (1.5,0.8) {$X_0^t$};
    \node at (3.5,0.8) {$U_{1,t}$};
    \node at (4.9,0.8) {$X_{1,t}$};
    \node at (6.3,0.8) {$Y_{1}^n$};
    \node at (8.3,0.8) {$U_{2}^n$};
    \node at (8.3,-0.2) {$X_{1}^n$};
    \node at (2.5,0.5) {$C_1$};
    \node at (7.3,0.5) {$C_2$};
\end{tikzpicture}
\label{fig:sim}

  \caption{Causal-encoding noncausal-decoding vector-valued Witsenhausen counterexample. 
  }
  \vspace{-0.6cm}
\end{figure}

Inspired by the two-point strategy, we propose a zero estimation cost (ZEC) approach for the causal-encoding setup based on the single-letter coordination coding result, where one aux. RV is designed to be Gaussian, and the other one is discrete representing the sign of the source state. The ZEC scheme requests the two aux. RVs to describe the controlled system state $X_1$ deterministically. Due to block coding, these two aux. RVs are subsequently revealed to the noncausal decoder, which facilitates the decoder to perfectly reconstruct the system state, thereby achieving zero estimation cost. Simulation results show that the \ac{zec} strategy significantly reduces the power required for zero-cost system state reconstruction at the decoder, offering a substantial improvement over existing single-shot methods. Next, we introduce a more general non-zero estimation cost (Non-ZEC) scheme by incorporating a test channel on top of the discrete aux. RV. This extended scheme reduces the necessary power cost by allowing a trade-off with estimation accuracy, which is shown numerically, to be a time-sharing operation between the original two-point scheme and our proposed ZEC strategy. By exploiting block-coding gain, our proposed approach strictly outperforms the two-point strategy and greatly enhances the power-estimation trade-off for Witsenhausen counterexample.

This paper is organized as follows: Section \ref{sec: system model} formulates the problem and recapitulates some foundational results. Section \ref{sec: lossless} introduces the ZEC scheme and its performance analysis. Section \ref{sec: lossy} extends this to the Non-ZEC scheme with numerical simulations. Lastly, a conclusion follows in Section \ref{sec: conclusion}.

\section{System Model}\label{sec: system model}


Let us consider the vector-valued Witsenhausen counterexample setup with causal source states and channel noises that are drawn independently according to the i.i.d. Gaussian distributions $X_0^n\sim\mathcal{N}(0, Q\mathbb I)$ and $Z_1^n\sim\mathcal{N}(0,N\mathbb I)$, for some $Q,N\in\mathbb R^+$, where $\mathbb I$ is the identity matrix, see Figure 1. We denote by $X_1$ the memoryless interim system state and $Y_1$ the output of the memoryless additive channel, generated by
\begin{flalign}
    &X_1 = X_0 + U_1  &\text{with }X_0\sim\mathcal{N}(0,Q),\label{X1 generation}\\
    &Y_1 = X_1 + Z_1 = X_0 + U_1 + Z_1 &\text{with }Z_1\sim\mathcal{N}(0,N).\label{Y1 generation}
\end{flalign}
We denote by $\mathcal{P}_{X_0} = \mathcal{N}(0,Q)$ the generative Gaussian probability distribution of the source, and by $\mathcal{P}_{X_1,Y_1|X_0,U_1}$ the channel probability distribution according to \eqref{X1 generation} and \eqref{Y1 generation}.

We define the control design for this setup, its induced cost functions, and the achievable cost pairs as follows:

\begin{definition}
    For $n\in\mathbb{N}$, a ``control design'' with causal encoder and noncausal decoder is a tuple of stochastic functions $c = (\{ f^{(t)}_{U_{1,t}|X_0^t}\}_{t=1}^n, g_{U_2^n|Y_1^n})$ defined by
    \begin{equation}
        f^{(t)}_{U_{1,t}|X_0^t}: \mathcal{X}_0^t \longrightarrow \mathcal{U}_1,\quad g_{U_2^n|Y_1^n}: \mathcal{Y}_1^n\longrightarrow \mathcal{U}_2^n,\label{eq: c-n control design}
    \end{equation}
    which induces a distribution over sequences of symbols:
    \begin{equation}
         \prod_{t=1}^n \mathcal{P}_{X_{0,t}} \prod_{t=1}^n f^{(t)}_{U_{1,t}|X_0^t}\prod_{t=1}^n \mathcal{P}_{X_{1,t},Y_{1,t}|X_{0,t},U_{1,t}} g_{U_2^n|Y_1^n}. \nonumber
    \end{equation}

\end{definition}

\begin{definition}\label{def: achievable cost}
    We define the two long-run cost functions $c_P(u_1^n) = \frac{1}{n}\sum_{t=1}^n (u_{1,t})^2$ and 
$c_S(x_1^n, u_2^n) = \frac{1}{n}\sum_{t=1}^n(x_{1,t}-u_{2,t})^2$. The pair of costs $(P,S)\in\mathbb{R}^2$ is said to be achievable if for all $\varepsilon>0$, there exists $\Bar{n}\in\mathbb N$ such that for all $n\geq \Bar{n}$, there exists a control design $c$ as in \eqref{eq: c-n control design} such that 
    \begin{equation}
        \mathbb E\Big[\big|P - c_P(U_1^n)\big| + \big|S - c_S(X_1^n, U_2^n)\big|\Big] \leq \varepsilon.\nonumber
    \end{equation}
\end{definition}

The following theorem is the single-letter characterization for the optimal cost region formed by all achievable cost pairs.

\begin{theorem}[\!\!\protect{\cite[Theorem II.3]{zhao2024coordination}}]\label{wits main theorem}
    The pair of Witsenhausen costs $(P,S)$ is achievable if and only if there exists a joint distribution over the random variables $(X_0, W_1, W_2, U_1, X_1, Y_1, U_2)$ that decomposes according to
    \begin{flalign}
\!     \mathcal{P}_{X_0}\mathcal{P}_{W_1}\mathcal{P}_{W_2|X_0,W_1}\mathcal{P}_{U_1|X_0,W_1}\mathcal{P}_{X_1, Y_1|X_0, U_1}\mathcal{P}_{U_2|W_1, W_2, Y_1},\label{eq: prob result}
    \end{flalign}
    such that
    \begin{align}
        &I(W_1, W_2; Y_1) - I(W_2; X_0 | W_1) \geq 0,\label{info result}\\
        &P = \mathbb{E}\sbrackets{U_1^2}, \quad \quad S = \mathbb{E}\sbrackets{(X_1 - U_2)^2},\nonumber
    \end{align}
    where $\mathcal{P}_{X_0}$ and $\mathcal{P}_{X_1, Y_1|X_0, U_1}$ are two given Gaussian distributions, and $W_1,W_2$ are two aux. RVs.
\end{theorem}

The two aux. RVs can be interpreted as follows: $W_1$ represents the independent codeword designed for the state-dependent channel with state $X_0$, consistent with the Shannon strategy \cite{shannonstrategy1958}. $W_2$ is correlated with both $X_0$ and $W_1$, acting as a description of these two symbols. Both $W_1$ and $W_2$ are made available to the noncausal decoder. This formulation explicitly captures the dual role of control in Witsenhausen counterexample. 

\begin{remark}
    The following Markov chains follow from the joint probability distribution \eqref{eq: prob result}:
    \begin{align}
        \left\{  
        \begin{aligned}
            &X_0\text{ is independent of }W_1,\\
            & U_1 -\!\!\!\!\minuso\!\!\!\!- (X_0, W_1) -\!\!\!\!\minuso\!\!\!\!- W_2,\\
            &(X_1, Y_1)-\!\!\!\!\minuso\!\!\!\!- (X_0, U_1)    -\!\!\!\!\minuso\!\!\!\!- (W_1, W_2),\\
            & U_2 -\!\!\!\!\minuso\!\!\!\!- (W_1, W_2, Y_1) -\!\!\!\!\minuso\!\!\!\!- (X_0, U_1, X_1).
        \end{aligned}
        \right.
        \label{markov result}
    \end{align}
\end{remark}
The first two Markov chains are consequences of causal encoding. The third Markov chain is related to the processing order of the Gaussian channel. The last Markov chain comes from non-causal decoding and symbol-wise reconstruction. 

In order to investigate the optimal achievable cost pairs, we focus on the lower boundary of the two-dimensional power-estimation cost region characterized in Theorem \ref{wits main theorem}: For a given power cost $P\geq 0$, we aim to determine the minimum estimation cost $S$ achievable at the decoder. Since the \ac{mmse} estimation provides the optimal decoding policy and is given by the conditional expectation, we have the following lemma:

\begin{lemma}
    Given a power cost parameter $P\geq 0$, the optimal estimation cost $S^*(P)$ is given by
        \begin{align}
            &S^*(P) = \inf_{\mathcal{P}\in\mathbb P(P)}\mathbb E\Big[\big(X_1 - \mathbb E\big[X_1\big|W_1,W_2,Y_1\big]\big)^2\Big],\label{eq: mmse}\\
            &\mathbb P(P) = \Bigl\{  (\mathcal{P}_{W_1},\mathcal{P}_{W_2|X_0, W_1},\mathcal{P}_{U_1|X_0, W_1}) \text{ s.t. }P = \mathbb E\big[U_1^2\big],\nonumber\\
            &\quad\quad\quad\quad I(W_1,W_2;Y_1) - I(W_2;X_0|W_1)\geq0\Bigr\}.\nonumber
        \end{align}
\end{lemma}

Next, we revisit Witsenhausen's two-point strategy. Our findings in \cite{zhao2024CDC} show it outperforms both the best affine \cite[Lemma III.3]{zhao2024CDC} and optimal joint Gaussian strategies \cite[Theorem III.4]{zhao2024CDC} for some values of $Q,N$.

\begin{theorem}[\protect{\cite[Theorem 2]{witsenhausen1968}}]
    For parameter $a\geq 0$, Witsenhausen's two-point strategy is given by
    \begin{align}
        U_1 = a\cdot \mathsf{sign}(X_0) - X_0. \label{eq: two-point}
    \end{align}
    The power and estimation costs are given by
    \begin{align}
        P_2(a) &= Q + a\brackets{a - 2\sqrt{\frac{2Q}{\pi}}},\label{eq: two-point cost P}\\
        S_2(a) &= a^2\sqrt{\frac{2\pi}{N}}\phi\brackets{\frac{a}{\sqrt{N}}}\int \frac{\phi\brackets{\frac{y_1}{\sqrt{N}}}}{\cosh{(\frac{ay_1}{N})}}dy_1,\label{eq: two-point cost S}
    \end{align}
    where $\phi(x) = \frac{1}{\sqrt{2\pi}}e^{-\frac{x^2}{2}}$ and the optimal receiver's strategy is given by $\mathbb E\big[X_1\big|Y_1 = y_1\big]=a\tanh{(\frac{ay_1}{N})}$.
\end{theorem}

The core idea of the two-point strategy is to cancel the continuous source state $X_0$, such that the encoder designates the system state $X_1=U_1+X_0=a\cdot \mathsf{sign}(X_0)$ to be binary, making it easier for the decoder to estimate. Inspired by this approach, in the next section, we introduce the ZEC scheme based on our coordination coding result in Theorem \ref{wits main theorem}. The ZEC method leverages the block-coding gain to reconstruct $X_1$ and achieves a zero estimation cost even at a significantly low power budget.
\vspace{0.1cm}

\section{The Zero Estimation Cost Scheme}\label{sec: lossless}

We propose the following design of aux. RVs involved in Theorem \ref{wits main theorem}, where $W_1$ is continuous Gaussian and $W_2$ is discrete binary\footnote{This approach is similar to the hybrid coding scheme \cite{skoglund2006hybrid}, which also combines digital and analog coding.}.  For given parameters $V_1\geq 0,a\geq 0$,
\begin{align}
\left|
\begin{aligned}
X_0 &\sim \mathcal{N}(0,Q),\\
W_1 &\sim \mathcal{N}(0,V_1),\\
W_2 &= a \cdot S,\text{ where } S = \mathsf{sign}(X_0),\\
U_1 &= W_1 + a\cdot S - X_0,\\
X_1 &= U_1 + X_0 = W_1 + a \cdot S = W_1 + W_2,\\
Y_1 &= X_1 + Z_1 = W_1 + W_2 + Z_1,\quad Z_1 \sim \mathcal{N}(0,N).
\end{aligned}
\right.
\label{eq:lossless system}
\end{align}

In this joint control-communication scheme, the control action $U_1$ subtracts the Gaussian source state $X_0$ as in \eqref{eq: two-point} and embeds the two aux. RVs $(W_1,W_2)$. This makes the system state $X_1$ offset the uncertainty of the source state and becomes a deterministic function of $(W_1,W_2)$. These two aux. RVs, due to block coding, can be communicated to the receiver efficiently. Hence, the noncausal decoder has access to sufficient information and outputs the MMSE estimation
\begin{align*}
    \mathbb E[X_1|W_1,W_2,Y_1] =  \mathbb E[W_1+W_2|W_1,W_2,Y_1] = W_1+W_2
\end{align*}
which perfectly reconstructs $X_1$, resulting in a zero estimation cost. Recall that, according to Definition \ref{def: achievable cost}, the estimation cost $S=0$ is achievable means that as the number of transmission $n\rightarrow\infty$, the averaged long-fun cost function $c_S(X_1^n, U_2^n)$ could be made as close to zero as possible.

Moreover, the power cost needed for this system is
\begin{align}
 P = \mathbb{E}[U_1^2]&= \mathbb E[(W_1 + a\cdot S - X_0)^2]\nonumber \\
    &= V_1 + \brackets{Q+a^2-2a\sqrt{\frac{2Q}{\pi}}}.  \nonumber
\end{align}
Therefore, for a fixed power cost $P\geq 0$, $V_1$ is uniquely determined by the parameter $a$ which needs to satisfy the following condition
\begin{align}
    V_1 = P - \brackets{Q+a^2-2a\sqrt{\frac{2Q}{\pi}}}\geq 0.\label{eq: coord2 power cost constraint}
\end{align}

The information constraint \eqref{info result} in bits becomes
\begin{align}
    &I(W_1,W_2;Y_1) - I(W_2;X_0|W_1)\nonumber\\
    &=h(Y_1)-h(Y_1|W_1,W_2)-h(W_2|W_1)+h(W_2|X_0,W_1)\nonumber\\
    &=h(Y_1) - \frac{1}{2}\log_2(2\pi eN) - 1\geq 0,\label{eq: coord2 info constraint}
\end{align}
where $h(Y_1)$ is calculated with regard to the Gaussian mixture distribution of the following form
\begin{align}
    f_{Y_1}(y)\!=\! \frac{1}{2\sqrt{(V_1+N)}}\!\sbrackets{\phi\brackets{\frac{y-a}{\sqrt{V_1+N}}}\!+\!\phi\brackets{\frac{y+a}{\sqrt{V_1+N}}}}\!,\label{eq: Y_1 distr}
\end{align}
where $\phi(x)=\frac{1}{\sqrt{2\pi}}\exp{\brackets{\frac{x^2}{2}}}$ is the standard Gaussian p.d.f.. Since there is no closed form for the entropy of Gaussian mixture distributions, methods discussed in \cite{huber2008entropy, kim2015entropy} can be employed for numerical simulation.

We denote the set of parameters $a\geq 0$ that satisfy the power cost constraint \eqref{eq: coord2 power cost constraint} and the information constraint \eqref{eq: coord2 info constraint} by
\begin{align}
    \mathcal{A}^0(P) &= \{a\geq 0: h(Y_1)- \frac{1}{2}\log_2(2\pi eN) \geq 1, \label{eq: admissible set}
\\
    &\quad\quad \text{and }
    V_1 = P - \brackets{Q + a^2 - 2a\sqrt{\frac{2Q}{\pi}}}\geq 0 \}.\nonumber \end{align}

Based on the above derivation, the optimal cost function for the ZEC system is summerized in the following theorem:
\begin{theorem}\label{thm: zec}
   Given the power cost $P\geq 0$, the \ac{mmse} estimation cost  for the ZEC coding scheme \eqref{eq:lossless system} is given by
    \begin{align}
        S_{\mathsf{ZEC}}(P) = 0, \text{ for  }P\geq P^*, \label{eq: cost function lossless}
    \end{align}
    where the value  
\begin{align}
    P^* = \min\{ P: \mathcal{A}^0(P) \neq\varnothing\}.\label{eq: P^*}
\end{align}

\end{theorem}

To satisfy the second nonnegative constraint of the admissible condition in \eqref{eq: admissible set}, the minimum required power cost $P^*$ for the ZEC scheme must satisfy
\begin{align*}
    P^*\geq \min_a \brackets{Q + a^2 - 2a\sqrt{\frac{2Q}{\pi}}} = Q\brackets{1-\frac{2}{\pi}}= P_2^{\min} ,
\end{align*}
where $ P_2^{\min}$ represents the minimum power budget required for the two-point strategy \eqref{eq: two-point cost P}. This indicates that the ZEC scheme achieves an improved estimation performance, at the expense of a higher power consumption than the original two-point strategy.

\begin{figure}[h]
        \centering

\definecolor{airforceblue}{rgb}{0.36,0.64,0.50}
\definecolor{antiquebrass}{rgb}{0.7,0.48,0.36}
\definecolor{alizarin}{rgb}{1.0, 0.6, 0.2}
\definecolor{amethyst}{rgb}{0.6,0.4,0.8}


\begin{tikzpicture}[scale=0.95]
\begin{axis}[
    xlabel={$P$},
    ylabel={$\mathsf{MMSE}$},
    legend pos=north east,
    axis lines=middle,
    axis line style={lightgray, line width=0.5pt},
    xmin=0, xmax=1.02,
    ymin=-0.0005, ymax=0.132,
    xtick={1},
    ytick=\empty,
    xlabel style={at={(ticklabel* cs:1)}, anchor=north west},
    ylabel style={at={(ticklabel* cs:1)}, anchor=south east},
]

\addplot[amethyst, line width=2.0pt, solid] table [col sep=comma, x index=0, y index=1] {fig/coord2.csv};
\addlegendentry{$S_{\mathsf{ZEC}}(P)$}

\addplot[alizarin, line width=1.9pt, dashed] table [col sep=comma, x index=0, y index=1] {fig/2P.csv};
\addlegendentry{$(P_2(a),S_2(a))$}

\addplot[airforceblue, line width=1.9pt, solid] table [col sep=comma, x index=0, y index=1] {fig/l.csv};
\addlegendentry{$S_{\ell}(P)$}

\addplot[antiquebrass, line width=1.9pt, dotted] table [col sep=comma, x index=0, y index=1] {fig/g.csv};
\addlegendentry{$S_{\mathsf{G}}(P)$}

\addplot[only marks, mark=*, mark options={scale=0.6}, color=black] coordinates {
    (0.3825, 0) (0.3634, 0.0376)(1,0.0)
};

\node[right, draw, rectangle, inner sep=1.7pt] at (axis cs:0.155, 0.0065) {$P^*=0.383$}; 
\node[above, draw, rectangle, inner sep=1.6pt] at (axis cs:0.182, 0.025) {$ P_2^{\min}=0.363$}; 

\end{axis}
\end{tikzpicture}

        \caption{Comparison of the four cost functions $S_{\mathsf{ZEC}}(P)$, $S_2(P)$, $S_\ell(P)$, and $S_{\mathsf{G}}(P)$ at $Q=1,N=0.15$. Our proposed scheme strictly outperforms the other strategies and achieves a zero-estimation-cost state reconstruction when $P\geq P^*= 0.383$.}
        \label{fig:Q=1,N=0.15}
    \end{figure}

To illustrate the performance of the ZEC scheme, we compare its cost function  $ S_{\mathsf{ZEC}}(P) $, with that of the original two-point strategy $S_2(P)$ given in \eqref{eq: two-point cost P}-\eqref{eq: two-point cost S}, the best affine strategy $ S_\ell(P)$ \cite[Lemma III.3]{zhao2024CDC}, and the optimal joint Gaussian strategy $S_\mathsf{G}(P)$ \cite[Theorem III.4]{zhao2024CDC} at $Q = 1, N = 0.15$ in Figure \ref{fig:Q=1,N=0.15}. As we can see, using only a slightly higher power cost than $P_2^{\min}=0.363$, the ZEC scheme can already achieve a zero-estimation-cost system state reconstruction when $P\geq P^*=0.383$. In contrast, all the other strategies, $S_2(P)$, $S_\ell(P)$, and $S_G(P)$, only achieve zero-estimation-cost reconstruction at a significantly bigger power cost of $P = Q = 1$.

    Figure \ref{fig:PvsN} illustrates how the minimum required power $P^*$ given in \eqref{eq: P^*} varies with different values of $N$ at $Q=1$. Notably, for small noise levels (e.g., $N\leq 0.07$), we observe that $P^*=P_2^{\min}=0.363$,  indicating that zero-cost estimation can be achieved without any additional power expenditure compared to the original two-point strategy. Furthermore, when $N\geq 0.3$, that is, scenarios when the two-point strategy no longer surpasses the optimal affine approach, our proposed scheme continues to achieve zero-cost estimation requiring only a power budget of $P\geq P^*= 0.501$. Note that this is only about half of the power budget needed by other strategies $S_\ell (P),S_{\mathsf{G}}(P),S_2(P)$ for achieving zero-cost system state estimation. However, when $N\geq 0.65$, $P^*\geq Q=1$, meaning that the ZEC scheme can no longer provide a zero-estimation-cost block-coding gain in high-noise regimes.
    
    \begin{figure}[ht]
        \centering

\definecolor{airforceblue}{rgb}{0.36,0.64,0.50}
\definecolor{antiquebrass}{rgb}{0.7,0.48,0.36}
\definecolor{alizarin}{rgb}{0.82,0.1,0.26}
\definecolor{amethyst}{rgb}{0.6,0.4,0.8}


\begin{tikzpicture}[scale=0.95]
\begin{axis}[
    xlabel={$N$},
    ylabel={$P^*(N)$},
    axis lines=middle,
    axis line style={lightgray, line width=0.5pt},
    xmin=0, xmax=0.705,
    ymin=0.3, ymax=1.005,
    xtick={0.1,0.2,0.3,0.4,0.5,0.6, 0.65},
    ytick={0.363, 1},
    scaled ticks=false, 
    tick label style={
        font=\small, 
        /pgf/number format/fixed, 
        /pgf/number format/precision=3, 
    },
    xlabel style={at={(ticklabel* cs:1)}, anchor=north west},
    ylabel style={at={(ticklabel* cs:1)}, anchor=south east},
    legend style={draw=none}, 
    legend entries={}, 
]

\addplot[airforceblue, line width=1.5pt, solid] table [col sep=comma, x index=0, y index=1] {fig/PvsN.csv};

\addplot[only marks, mark=*, mark options={scale=0.5}, color=black] coordinates {
  (0, 0.363)   (0.1, 0.367) (0.2, 0.411) (0.3, 0.501) (0.4, 0.624) (0.5, 0.768) (0.6, 0.928) (0.645, 1) (0, 1)
};


\node[above] at (axis cs:0.07, 0.377) {\tiny $P^*=0.367$}; 
\node[above] at (axis cs:0.17, 0.431) {\tiny $P^*=0.411$}; 
\node[above] at (axis cs:0.25, 0.521) {\tiny $P^*=0.501$}; 
\node[above] at (axis cs:0.35, 0.644) {\tiny $P^*=0.624$}; 
\node[above] at (axis cs:0.45, 0.788) {\tiny $P^*=0.768$}; 
\node[above] at (axis cs:0.55, 0.948) {\tiny $P^*=0.928$}; 

\draw [dotted] (axis cs:0.1, 0.367) -- (axis cs:0.1, 0.0);
\draw [dotted] (axis cs:0.2, 0.411) -- (axis cs:0.2, 0.0);
\draw [dotted] (axis cs:0.3, 0.501) -- (axis cs:0.3, 0.0);
\draw [dotted] (axis cs:0.4, 0.624) -- (axis cs:0.4, 0.0);
\draw [dotted] (axis cs:0.5, 0.768) -- (axis cs:0.5, 0.0);
\draw [dotted] (axis cs:0.6, 0.928) -- (axis cs:0.6, 0.0);
\draw [dotted] (axis cs:0.647, 1.0) -- (axis cs:0.65, 0.0);

\draw [dotted] (axis cs:0.65, 1.0) -- (axis cs:0, 1.0);

\end{axis}
\end{tikzpicture}

        \caption{Variation of $P^*$ as a function of the noise level $N$ when $Q=1$. $P^*=P_2^{\min}=0.363$ when $N$ is small, and increases to $P^*=Q=1$ for $N\geq 0.65$.}
        \label{fig:PvsN}
    \end{figure}

Next, we extend the ZEC strategy to a more general Non-ZEC scheme. This extension introduces a trade-off between estimation accuracy and power cost, enabling a power cost reduction.

\vspace{0.5cm}

\section{The Non-Zero Estimation Cost Scheme}\label{sec: lossy}

In this section, we apply a test channel between the aux. RV $W_2$ and the source state $X_0$ with a cross-over probability $\gamma$. Given nonnegative parameters $V_1,a,\gamma$, we consider
\begin{align}
\left|
\begin{aligned}
    &X_0\sim\mathcal{N}(0,Q),\\
    & W_1\sim\mathcal{N}(0, V_1),\\
    & W_2=\left\{
          \begin{aligned}
              & a\cdot S &\text{with probability }1-\gamma,\\
              & -a\cdot S&\text{with probability }\gamma,
          \end{aligned}   
        \right.\\
        &\quad\quad\quad  \text{ where } S=\mathsf{sign}(X_0),\\
    & U_1 = W_1 + a\cdot S - X_0,\\
    & X_1 = U_1+X_0 = W_1 + a\cdot S, \\
    & Y_1 = X_1 + Z_1 = W_1 + a\cdot S+ Z_1,\quad Z_1\sim\mathcal{N}(0,N).
\end{aligned}
\right. 
\label{eq: lossy system}
\end{align}

The design of $W_2$ as a binary RV dependent on $X_0$ rather than a deterministic variable introduces additional randomness, which affects the estimation precision of $X_1$ from $(W_1,W_2,Y_1)$. Note that, if $\gamma=0$ or $1$, the above system degrades to a deterministic relation, where $X_1=W_1+W_2$ or $X_1=W_1-W_2$, respectively. In either case, the estimation cost is zero, hence we recover the ZEC scheme. Furthermore, it can be easily shown that the \ac{mmse} remains the same for $\gamma$ and $1-\gamma$. Therefore, we can restrict our analysis to $\gamma\in[0,0.5]$, without loss of generality.

To simplify the presentation of the estimation cost result presented in Theorem \ref{thm: coord_2_testch}, we define the following quantities:
\begin{align*}
     &G_0(w_1) =\frac{1}{\sqrt{V_1}}\phi\brackets{\frac{w_1}{\sqrt{V_1}}},\\
     &G_1(w_1,y_1) = \frac{1}{\sqrt{N}}\phi\brackets{\frac{y_1-(w_1+a)}{\sqrt{N}}},\\
     &G_2(w_1,y_1) = \frac{1}{\sqrt{N}}\phi\brackets{\frac{y_1-(w_1-a)}{\sqrt{N}}},\\
     &I(w_1,y_1)\!=\!G_0\!\sbrackets{\frac{ [(1-\gamma)G_1 - \gamma G_2]^2}{(1-\gamma)G_1 + \gamma G_2}\! +\! \frac{ [\gamma G_1 - (1-\gamma) G_2]^2}{\gamma G_1 + (1-\gamma) G_2}}.
\end{align*}
Similarly, we define the set of admissible parameters
\begin{align}
    \!\mathcal{A}^\gamma(P)&\! =\! \{a\geq 0:h(Y_1)+H_2(\gamma) - h(Y_1|W_1,W_2)\geq 1,\nonumber \\
    &\quad \text{and } V_1 = P-\brackets{Q+a^2-2a\sqrt{\frac{2Q}{\pi}}}\geq 0\}, \label{eq: admissible set lossy}
\end{align}
where $h(Y_1)$ is the entropy calculated from the distribution defined in \eqref{eq: Y_1 distr}, $H_2(\gamma)=-\gamma \log_2(\gamma)-(1-\gamma)\log_2(1-\gamma)$ is a binary entropy, and $h(Y_1|W_1,W_2)$ is the following Gaussian mixture differential entropy
\begin{align*}
    h\brackets{(1-\gamma) \frac{1}{\sqrt{N}}\phi\brackets{\frac{y-a}{\sqrt{N}}} + \gamma \frac{1}{\sqrt{N}}\phi\brackets{\frac{y+a}{\sqrt{N}}}}.
\end{align*}

By setting $\gamma=0$ (and also $\gamma=1$), \eqref{eq: admissible set lossy} recovers the admissible parameter set $\mathcal{A}^0(P)$ given in \eqref{eq: admissible set} of the ZEC scheme. Therefore,
\begin{align}
    \mathcal{A}^0 (P) \subseteq \mathcal{A}^\gamma (P),
\end{align}
indicating that the Non-ZEC scheme permits lower power costs compared to the ZEC scheme.

The full proof using the above quantities is in the appendix. Additionally, the derivation of the ZEC cost function sketched in equations \eqref{eq: coord2 power cost constraint} - \eqref{eq: admissible set} can also be seen as a special case of the derivation of the Non-ZEC scheme by setting $\gamma = 0$.

 \begin{theorem}\label{thm: coord_2_testch}
    Given $P\geq P_2^{\min}$, the optimal estimation cost induced by the Non-ZEC scheme \eqref{eq: lossy system} is given by
    \begin{align}
         S_{\mathsf{Non}\text{-}\mathsf{ZEC}}(P)= \min_{a\in\mathcal{A}^\gamma(P),\gamma\in[0,0.5]}\quad\big\{F(a,\gamma,P)\big\},\label{eq: cost func lossy}
         \end{align}
where,
         \begin{align}
         F(a,\gamma,P)=a^2 -\frac{a^2}{2}\iint I(w_1,y_1) dw_1dy_1. \label{eq: F(a,gamma,P)}
    \end{align}
    
\end{theorem}

In this theorem, $F(a,\gamma,P)$ represents the achievable estimation cost for a given pair $(a,\gamma)$ at the power budget $P\geq P_2^{\min}$, and we can minimize $F$ over all admissible parameters $a\in\mathcal{A}^\gamma(P),\gamma\in[0,0.5]$ to get the optimized result \eqref{eq: cost func lossy}. Moreover, by plugging in $\gamma=0$, $S_{\mathsf{Non}\text{-}\mathsf{ZEC}}(P)$ boils down to $S_{\mathsf{ZEC}}(P) = 0$ when $P\geq P^*$ given in \eqref{eq: P^*}.

\begin{figure}[h]
        \centering

\definecolor{airforceblue}{rgb}{0.36,0.64,0.50}
\definecolor{antiquebrass}{rgb}{0.7,0.48,0.36}
\definecolor{alizarin}{rgb}{1.0, 0.6, 0.2}
\definecolor{amethyst}{rgb}{0.6,0.4,0.8}
\colorlet{amethystLight}{amethyst!50!white}


    \begin{tikzpicture}[scale=0.92]
\begin{axis}[
            name=plot1, 
            at={(0,0)},
            width=5cm, 
            xlabel={$P$},
            ylabel={$\mathsf{MMSE}$},
            legend pos=north east,
            axis lines=middle,
            axis line style={lightgray, line width=0.5pt},
            xmin=0, xmax=1.02,
            ymin=-0.0005, ymax=0.132,
            xtick={0.383},
            ytick=\empty,
            xlabel style={at={(ticklabel* cs:1)}, anchor=north},
            ylabel style={at={(ticklabel* cs:1)}, anchor=south},
            tick label style={
                /pgf/number format/fixed,
                /pgf/number format/precision=3,
            },
        ]
        \addplot[amethyst, line width=0.95pt, solid] table [col sep=comma, x index=0, y index=1] {fig/coord2.csv};
        \addlegendentry{$F(a,0,P)$}
        \addplot[alizarin, line width=1.5pt, dashed] table [col sep=comma, x index=0, y index=1] {fig/2P.csv};

\addplot[only marks, mark=*, mark options={scale=0.6}, color=black] coordinates {
    (0.383,0.0)
};
\node[below] at (axis cs: 0.5,-0.5) {Label for Plot 1};
    
    \end{axis}

\begin{axis}[
    name=plot2,
    at={($(plot1.south)+(3cm,0)$)},
    width=5cm,
    xlabel={$P$},
    ylabel={$\mathsf{MMSE}$},
    legend pos=north east,
    axis lines=middle,
    axis line style={lightgray, line width=0.5pt},
    xmin=0, xmax=1.02,
    ymin=-0.0005, ymax=0.132,
    xtick={0.376},
    ytick={0.012},
    xlabel style={at={(ticklabel* cs:1)}, anchor=north},
    ylabel style={at={(ticklabel* cs:1)}, anchor=south},
    tick label style={
                /pgf/number format/fixed,
                /pgf/number format/precision=3,
            },
]
\addplot[amethyst, line width=0.8pt, solid] table [col sep=comma, x index=0, y index=1] {fig/S05.csv};
\addlegendentry{$F(a,0.05,P)$};

\addplot[alizarin, line width=1.5pt, dashed] table [col sep=comma, x index=0, y index=1] {fig/2P.csv};

\addplot[only marks, mark=*, mark options={scale=0.6}, color=black] coordinates {
    (0.376,0.012)
};
\draw [dotted] (axis cs:0.376,0.012) -- (axis cs:0.376, 0.0);
\draw [dotted] (axis cs:0.376,0.012) -- (axis cs:0.0, 0.012);
\end{axis}

        



        
        \begin{axis}[
            name=plot3, 
            at={($(plot1.west)+(0cm,-5.5cm)$)}, 
            width=5cm, 
            xlabel={$P$},
            ylabel={$\mathsf{MMSE}$},
            legend pos=north east,
            axis lines=middle,
            axis line style={lightgray, line width=0.5pt},
            xmin=0, xmax=1.02,
            ymin=-0.0005, ymax=0.132,
            xtick={0.3725},
            ytick={0.018},
            xlabel style={at={(ticklabel* cs:1)}, anchor=north},
            ylabel style={at={(ticklabel* cs:1)}, anchor=south},
            tick label style={
                /pgf/number format/fixed,
                /pgf/number format/precision=3,
            },
        ]
        \addplot[amethyst, line width=0.8pt, solid] table [col sep=comma, x index=0, y index=1] {fig/S1.csv};
        \addlegendentry{$F(a,0.1,P)$}
        \addplot[alizarin, line width=1.5pt, dashed] table [col sep=comma, x index=0, y index=1] {fig/2P.csv};

\addplot[only marks, mark=*, mark options={scale=0.6}, color=black] coordinates {
    (0.3725,0.018)
};
        \draw [dotted] (axis cs:0.3725,0.018) -- (axis cs:0.3725, 0.0);
\draw [dotted] (axis cs:0.3725,0.018) -- (axis cs:0.0, 0.018);

    \end{axis}

        \begin{axis}[
            name=plot4, 
            at={($(plot3.west)+(4.7cm,0cm)$)}, 
            anchor=west,
            width=5cm, 
            xlabel={$P$},
            ylabel={$\mathsf{MMSE}$},
            legend pos=north east,
            axis lines=middle,
            axis line style={lightgray, line width=0.5pt},
            xmin=0, xmax=1.02,
            ymin=-0.0005, ymax=0.132,
            xtick={0.363},
            ytick={0.036},
            xlabel style={at={(ticklabel* cs:1)}, anchor=north},
            ylabel style={at={(ticklabel* cs:1)}, anchor=south},
            tick label style={
                /pgf/number format/fixed,
                /pgf/number format/precision=3,
            },
        ]
        \addplot[amethyst, line width=0.8pt, solid] table [col sep=comma, x index=0, y index=1] {fig/S5.csv};
        \addlegendentry{$F(a,0.5,P)$}
        \addplot[alizarin, line width=1.5pt, dashed] table [col sep=comma, x index=0, y index=1] {fig/2P.csv};
        \addplot[only marks, mark=*, mark options={scale=0.6}, color=black] coordinates {
    (0.363,0.036)
};
        \draw [dotted] (axis cs:0.363,0.036) -- (axis cs:0.363, 0.0);
        \draw [dotted] (axis cs:0.363,0.036) -- (axis cs:0.0, 0.036);
        \end{axis}

    \end{tikzpicture}

        \caption{Evolution of $F(a,\gamma,P)$ with different values of $\gamma=0,0.05, 0.1,0.5$. When $\gamma=0$, $F(a,0,P) = S_{\mathsf{ZEC}}(P)$ in \eqref{eq: cost function lossless}, and when $\gamma=0.5$, the lower boundary of $F(a,0.5,P)$ recovers that of the two-point strategy (yellow dashed curve).}
        \label{fig: diff gamma}
    \end{figure}

To analyze the performance of the Non-ZEC scheme, we examine how different values of $\gamma\in[0,0.5]$ affect its achievable cost region. Figure \ref{fig: diff gamma} illustrates the function $F(a,\gamma,P)$ in \eqref{eq: F(a,gamma,P)} by plotting \textit{all} the admissible points $a\in\mathcal{A}^\gamma(P)$ at each $P\geq P_2^{\min}$, resulting in a 2-dimensional region, for fixed parameters of $\gamma\in \{0,0.01,0.1,0.5\}$ compared to the original two-point strategy at $Q=1,N=0.15$.

As shown in Figure \ref{fig: diff gamma}, $F(a,0,P)$ aligns with the ZEC cost function $S_{\mathsf{ZEC}}(P)$ in \eqref{eq: cost function lossless}, as expected. Gradually increasing  $\gamma$, which introduces more randomness in $W_2$, allows for a reduction in power cost at the expense of sacrificing estimation accuracy. Ultimately, when  $\gamma=0.5$, which means $W_2$ provides no information about the source $X_0$ at all, the required power budget reaches its minimum, and the lower boundary of the achievable cost $F(a,0.5,P)$ aligns with that of the original two-point strategy.

\begin{figure}[t]
        \centering      

\definecolor{airforceblue}{rgb}{0.36,0.64,0.50}
\definecolor{antiquebrass}{rgb}{0.7,0.48,0.36}
\definecolor{alizarin}{rgb}{1.0, 0.6, 0.2}
\definecolor{amethyst}{rgb}{0.6,0.4,0.8}
\definecolor{lightgrey}{rgb}{0.7, 0.7, 0.7}


\begin{tikzpicture}[scale=0.95]
\begin{axis}[
    xlabel={$P$},
    ylabel={$\mathsf{MMSE}$},
    legend pos=north east,
    axis lines=middle,
    axis line style={lightgray, line width=0.5pt},
    xmin=0, xmax=1.02,
    ymin=-0.0007, ymax=0.132,
    xtick={0.363, 0.383, 1},
    xticklabels={{\hspace{-25pt}0.363}, {\hspace{25pt}0.383}, {1}},
    ytick=\empty,
    xlabel style={at={(ticklabel* cs:1)}, anchor=north west},
    ylabel style={at={(ticklabel* cs:1)}, anchor=south east},
    tick label style={
                /pgf/number format/fixed,
                /pgf/number format/precision=3,
            },
]

\addplot[amethyst, line width=2.0pt, solid] table [col sep=comma, x index=0, y index=1] {fig/coord2lossy.csv};
\addlegendentry{$S_{\mathsf{Non}\text{-}\mathsf{ZEC}}(P)$}

\addplot[alizarin, line width=2.1pt, dashed] table [col sep=comma, x index=0, y index=1] {fig/2P.csv};
\addlegendentry{$(P_2(a),S_2(a))$}

\addplot[airforceblue, line width=2.0pt, solid] table [col sep=comma, x index=0, y index=1] {fig/l.csv};
\addlegendentry{$S_{\ell}(P)$}

\addplot[antiquebrass, line width=2.0pt, dotted] table [col sep=comma, x index=0, y index=1] {fig/g.csv};
\addlegendentry{$S_{\mathsf{G}}(P)$}

\addplot[lightgrey, line width=2.0pt, loosely dashed] table [col sep=comma, x index=0, y index=1] {fig/timesharing.csv};
\addlegendentry{$S_{\mathsf{t}\text{-}\mathsf{s}}(P)$}

\addplot[only marks, mark=*, mark options={scale=0.6}, color=black] coordinates {
    (0.3825, 0) (0.3634, 0.0376)(1,0.0)
};

\draw [dotted, line width=0.8pt] (axis cs:0.3634, 0.0376) -- (axis cs:0.3634, 0.0);

\end{axis}
\end{tikzpicture}

        \caption{Comparison of the four closed-form cost functions $S_{\mathsf{Non}\text{-}\mathsf{ZEC}}(P)$, $S_2(P)$, $S_\ell(P)$, $S_{\mathsf{G}}(P)$, and the induced time-sharing cost $S_{\mathsf{t}\text{-}\mathsf{s}}(P)$ at $Q=1,N=0.15$. }
        \label{fig:Q=1,N=0.15,lossy}
    \end{figure}

In Figure \ref{fig:Q=1,N=0.15,lossy}, We plot the optimized achievable estimation cost function $S_{\mathsf{Non}\text{-}\mathsf{ZEC}}(P)$ in \eqref{eq: cost func lossy} and compare it with the other three cost functions $S_2(P), S_\ell(P)$ and $S_{\mathsf{G}}(P)$. The numerical results indicate that the Non-ZEC scheme functions effectively as a time-sharing mechanism between the ZEC scheme and the original two-point strategy. Moreover, $S_{\mathsf{t}\text{-}\mathsf{s}}(P)$ represents the estimation cost resulting from the time-sharing operation between the optimal Gaussian strategy $S_{\mathsf{G}}(P)$ and the ZEC strategy $S_{\mathsf{ZEC}}(P)$, which in this scenario, is shown to be superior to the Non-ZEC scheme.

\section{Conclusion}\label{sec: conclusion}
Our proposed joint control-communication schemes improve the overall power-estimation performance of Witsenhausen counterexample. In particular, the ZEC scheme significantly reduces the minimum power budget required for zero-estimation-cost system state reconstruction at the decoder from $P=Q$ to $P^*$. However, it remains unknown whether $P^*$ is the universal minimum power to achieve zero-cost estimation in the causal-encoding noncausal-decoding scenario. Furthermore, the current numerical results show that the Non-ZEC scheme is outperformed by the time-sharing mechanism between the optimal joint Gaussian strategy and the ZEC scheme. Whether there exist scenarios such that the Non-ZEC scheme can outperform this time-sharing mechanism remains unknown.

\bibliographystyle{ieeetr}
\bibliography{IEEEabrv,main}

\newpage
\onecolumn
\section*{Appendix}
\begin{proof}[Full Derivation for Theorem \ref{thm: coord_2_testch}]

Since the Non-ZEC coordination-coding scheme described in \eqref{eq: lossy system} involves both continuous and discrete RVs, our analysis must incorporate both the Lebesgue measure $\lambda$ (for continuous RVs) and the counting measure $\mu$ (for discrete RVs). The corresponding information-theoretic quantities are defined using the Radon-Nikodym derivative, which generalizes the concept of density with respect to a base measure. This approach enables us to define entropy for mixed discrete-continuous RVs in a way that preserves consistency with both discrete entropy and differential entropy; see \cite{pinsker1964information} for further details.

Using this framework, the information constraint can be reformulated as
    \begin{align}
        &I(W_1,W_2;Y_1) - I(W_2;X_0|W_1)\nonumber\\
    &= h(Y_1)+h(W_2|X_0,W_1) - h(Y_1|W_1,W_2)-h(W_2)\label{app info constraint 1}\\
    &= h(Y_1)+h(W|X_0,W_1) - h(Y_1|W_1,W_2)-h(W), \label{app info constraint 2}
    \end{align}
   where step \eqref{app info constraint 1} follows from the chain rule of mutual information, and in step \eqref{app info constraint 2}, we use the substitution:
    \begin{align*}
        W=W_2/a\sim\left\{
          \begin{aligned}
              & S &\text{with probability }1-\gamma,\\
              & -S&\text{with probability }\gamma,
          \end{aligned}   
        \right.
    \end{align*}
   with $S = \mathsf{sign}(X_0)$, a discrete random variable. Furthermore, by the scaling property of differential entropy, i.e., $h(cX) = h(X) + \log |c|$ for a constant $c \ne 0$, the constant $\log|a|$ that appears in both $h(W_2|X_0, W_1)$ and $h(W_2)$ cancels out, which justifies the equivalence between \eqref{app info constraint 1} and \eqref{app info constraint 2}.
 
Next, we examine each entropy term involved in \eqref{app info constraint 2} respectively
\begin{itemize}
    \item $h(Y_1):$ Since $ Y_1 = W_1 + a\cdot S + Z_1$ and $W_1\indep Z_1$ are both Gaussian distributed, the conditional distribution is
\begin{align*}
    f_{Y_1|S}(y|s ) = \frac{1}{\sqrt{2\pi(V_1+N)}}\exp{\brackets{-\frac{(y-a\cdot s)^2}{2(V_1+N)}}}.
\end{align*}
Moreover, since $\mathbb P(S=-1) = \mathbb P(S=+1) = \frac{1}{2}$, we have 
\begin{align}
    f_{Y_1}(y) &= \sum_{s\in\{-1,+1\}}f_{Y_1|S}(y|s )\mathbb P(s)\nonumber\\
    &=\frac{1}{2\sqrt{(V_1+N)}}\brackets{\phi\brackets{\frac{y-a}{\sqrt{V_1+N}}}+ \phi\brackets{\frac{y+a}{\sqrt{V_1+N}}}},\label{eq: term 1}
\end{align}
according to which we can calculate $h(Y_1)$.
\item $h(W|X_0,W_1)$: Since $W$ is independent of $W_1=w_1$ once $X_0=x_0$ is given, we have
\begin{align*}
    W|X_0=x_0,W_1=w_1\sim \left\{
          \begin{aligned}
              & s &\text{with probability }1-\gamma,\\
              & -s&\text{with probability }\gamma,
          \end{aligned}   
        \right.
\end{align*}
where $s = \mathsf{sign}(x_0)$. This defines a binary discrete distribution. Thus, $h(W|X_0,W_1)$ is a binary entropy that takes the following form
\begin{align}
    h(W|X_0,W_1) = H_2(\gamma)
    = -\sbrackets{(1-\gamma)\log_2(1-\gamma) + \gamma \log_2\gamma}.\label{eq: term 2}
\end{align}
\item $h(Y_1|W_1,W_2)$: Because 
\begin{align*}
    h(Y_1|W_1,W_2)
    &= h(W_1 + a\cdot S+Z_1|W_1,W_2)\\
    &=h( a\cdot S+Z_1|W_2).
\end{align*}
The conditional distribution takes the following form
\begin{align*}
    &a\cdot S+Z_1|W_2 = a\sim \left\{
          \begin{aligned}
              & a+Z_1 &\text{with probability }1-\gamma,\\
              & -a+Z_1&\text{with probability }\gamma.
          \end{aligned}   
        \right.
\end{align*}
Moreover, since $Z_1\sim\mathcal{N}(0,N)$ is the independent additive noise, 
\begin{align*}
    f(a\cdot S+Z_1|W_2=a)
    =(1-\gamma) \frac{1}{\sqrt{N}}\phi\brackets{\frac{y-a}{\sqrt{N}}} + \gamma \frac{1}{\sqrt{N}}\phi\brackets{\frac{y+a}{\sqrt{N}}},
\end{align*}
is also a Gaussian mixture distribution.

And similarly,
\begin{align*}
    f(a\cdot S+Z_1|W_2=-a)=\gamma \frac{1}{\sqrt{N}}\phi\brackets{\frac{y-a}{\sqrt{N}}} + (1-\gamma) \frac{1}{\sqrt{N}}\phi\brackets{\frac{y+a}{\sqrt{N}}}.
\end{align*}
We can see that the above two Gaussian mixture distributions are just swapped mixture weights and the shape of one is just “mirrored” left-to-right of the other one. Hence, the differential entropy (which is shift-invariant in the sense of mixing) remains the same. Therefore,
\begin{align}
    &h(a\cdot sign(X_0)+Z_1|W_2)\nonumber\\
    &=\frac{1}{2}h(a\cdot sign(X_0)+Z_1|W_2=a) + \frac{1}{2}h(a\cdot sign(X_0)+Z_1|W_2=-a) \nonumber\\
    &=h(a\cdot sign(X_0)+Z_1|W_2=a)\nonumber\\
    &=h\brackets{(1-\gamma) \frac{1}{\sqrt{N}}\phi\brackets{\frac{y-a}{\sqrt{N}}} + \gamma \frac{1}{\sqrt{N}}\phi\brackets{\frac{y+a}{\sqrt{N}}}}.\label{eq: term 3}
\end{align}

\item $h(W)$: Since the marginal $\mathbb P(W=1) = \mathbb P(W=-1) = \frac{1}{2}$, we obtain that
\begin{align}
    h(W)=1.\label{eq: term 4}
\end{align}

\end{itemize}

Given the above results \eqref{eq: term 1} - \eqref{eq: term 4} for calculating the information constraint, together with the power constraint, same as in \eqref{eq: coord2 power cost constraint}, the admissible set is given by
\begin{align*}
    \!\mathcal{A}^\gamma(P)&\! =\! \{a\geq 0:h(Y_1)+H_2(\gamma) - h(Y_1|W_1,W_2)\geq 1,\nonumber \\
    &\quad \text{and } V_1 = P-\brackets{Q+a^2-2a\sqrt{\frac{2Q}{\pi}}}\geq 0\}, 
\end{align*}
as provided in \eqref{eq: admissible set lossy}.

Next, given $\gamma\in[0,0.5],P\geq 0$, and for a feasible parameter $a\in\mathcal{A}^\gamma (P)$, we are interested in deriving the \ac{mmse} estimator of $X_1$ given $W_1=w_1,W_2=w_2,Y_1=y_1$, namely,
\begin{align*}
    \mathbb E[X_1|w_1,w_2,y_1] = \int x_1 f(x_1|w_1,w_2,y_1) dx_1.
\end{align*}
Hence, we need the closed-form expression of the following conditional probability distribution:
\begin{align*}
    \mathbb P(x_1|w_1,w_2,y_1)&= \mathbb P(w_1+a\cdot s|w_1,w_2,y_1)\\
    &= \mathbb P(s|w_1,w_2,y_1),
\end{align*}
which reduces to a two-point distribution. Since
\begin{align}
    \mathbb P(s=1|w_1,w_2,y_1)
    &=\frac{\mathbb P(s=1,w_2,y_1|w_1)}{\mathbb P(w_2,y_1|w_1)}\nonumber\\
     &=\frac{\mathbb P(w_2,y_1|w_1,s=1)\mathbb P(s=1|w_1)}{\sum_{s'\in\{-1,1\}} \mathbb P(w_2,y_1|w_1,s')\mathbb P(s'|w_1)}\nonumber\\
    &=\frac{\mathbb P(w_2,y_1|w_1,s=1)\mathbb P(s=1)}{\sum_{s'\in\{-1,1\}} \mathbb P(w_2,y_1|w_1,s')\mathbb P(s')}\nonumber\\
    &=\frac{\frac{1}{2}\mathbb P(w_2,y_1|w_1,s=1)}{\sum_{s'\in\{-1,1\}}\frac{1}{2} \mathbb P(w_2,y_1|w_1,s')}\nonumber\\
    &=\frac{\mathbb P(w_2,y_1|w_1,s=1)}{\sum_{s'\in\{-1,1\}}\mathbb P(w_2,y_1|w_1,s')}\nonumber\\
    &= \frac{\mathbb P(y_1|w_1,s=1)\cdot\mathbb P(w_2|s=1,w_1,y_1)}{\sum_{s'\in\{-1,1\}}\mathbb P(w_2,y_1|w_1,s')}   \label{eq: prob coord-2 text channel}
\end{align}
Since $Y_1 = W_1+a\cdot S + Z_1$, the term $\mathbb P(y_1|w_1,s=1)$ above becomes a Gaussian distribution
\begin{align*}
    \mathbb P(y_1|w_1,s=1) = \frac{1}{\sqrt{N}}\phi\brackets{\frac{y_1-(w_1+a)}{\sqrt{N}}}\triangleq G_1.
\end{align*}
 Moreover, because of the Markov chain $W_2-\!\!\!\!\minuso\!\!\!\!- S    -\!\!\!\!\minuso\!\!\!\!- (W_1, Y_1)$, we obtain
\begin{align*}
   \mathbb P(w_2|s=1,w_1,y_1) = \mathbb P(w_2|s=1) .
\end{align*}
Therefore, \eqref{eq: prob coord-2 text channel} becomes
\begin{align*}
   \mathbb P(s=1|w_1,w_2,y_1)
    =\frac{\mathbb P(w_2|s=1)G_1}{\mathbb P(w_2|s=1)G_1 + \mathbb P(w_2|s=-1)G_2}.
\end{align*}
where $G_2\triangleq \frac{1}{\sqrt{N}}\phi\brackets{\frac{y_1-(w_1-a)}{\sqrt{N}}}$.

Therefore, 
\begin{align*}
    \mathbb P(s=-1|w_1,w_2,y_1)
    &= 1-\frac{\mathbb P(w_2|s=1)G_1}{\mathbb P(w_2|s=1)G_1 + \mathbb P(w_2|s=-1)G_2}\\
    &= \frac{\mathbb P(w_2|s=-1)G_2}{\mathbb P(w_2|s=1)G_1 + \mathbb P(w_2|s=-1)G_2}.
\end{align*}
This means, 
\begin{align*}
    &\mathbb P(x_1=w_1+a|w_1,w_2,y_1)
    = \frac{\mathbb P(w_2|s=1)G_1}{\mathbb P(w_2|s=1)G_1 + \mathbb P(w_2|s=-1)G_2},\\
    &\mathbb P(x_1=w_1-a|w_1,w_2,y_1) 
    = \frac{\mathbb P(w_2|s=-1)G_2}{\mathbb P(w_2|s=1)G_1 + \mathbb P(w_2|s=-1)G_2}.\label{eq: distr x_1=w_1-a}
\end{align*}
Hence, when $w_2=a$, the MMSE estimator is
\begin{align*}
    &\mathbb E[X_1|W_1=w_1,W_2=a,Y_1=y_1] \\
    &=w_1 + \frac{ \mathbb P(w_2=a|s=1)G_1 - \mathbb P(w_2=a|s=-1) G_2}{ \mathbb P(w_2=a|s=1)G_1 +  \mathbb P(w_2=a|s=-1) G_2}\cdot a\\
    &=w_1 + \frac{ (1-\gamma)G_1 - \gamma G_2}{(1-\gamma)G_1 + \gamma G_2}\cdot a.
\end{align*}
And on the other hand, when $w_2=-a$,
\begin{align*}
    &\mathbb E[X_1|W_1=w_1,W_2=-a,Y_1=y_1] \\
    &=w_1 + \frac{ \mathbb P(w_2=-a|s=1)G_1 - \mathbb P(w_2=-a|s=-1)G_2}{ \mathbb P(w_2=-a|s=1)G_1 +  \mathbb P(w_2=-a|s=-1) G_2}\cdot a\\
    &=w_1 + \frac{ \gamma G_1 - (1-\gamma) G_2}{\gamma G_1 + (1-\gamma) G_2}\cdot a.
\end{align*}
And the joint distribution of $(w_1,w_2,y_1)$ when $w_2=a$ is
\begin{align*}
    \mathbb P(w_1,w_2=a,y_1) &=f(w_1)\cdot \mathbb P(a,y_1|w_1) \\
    &=f(w_1)\brackets{\sum_{s'\in\{-1,1\}} \mathbb P(a,y_1|w_1,s')\mathbb P(s')}\\
    &=\frac{f(w_1)}{2}\brackets{\sum_{s'\in\{-1,1\}} \mathbb P(y_1|w_1,s')\cdot\mathbb P(a|s')}\\
    &=\frac{G_0}{2}\sbrackets{(1-\gamma)G_1 + \gamma G_2} ,
\end{align*}
where $G_0\triangleq \frac{1}{\sqrt{V_1}}\phi\brackets{\frac{w_1}{\sqrt{V_1}}}$. Similarly, if we plug in $w_2=-a$, we get
\begin{align*}
     \mathbb P(w_1,w_2=-a,y_1) =\frac{G_0}{2}\sbrackets{\gamma G_1 + (1-\gamma) G_2} .
\end{align*}
Next, we calculate the expected squared MMSE estimation in the following way:
\begin{align}
    &\mathbb{E}\sbrackets{\left(\mathbb{E}\sbrackets{X_1|W_1,W_2,Y_1}\right)^2} \\
    &=\iiint (\mathbb E[X_1|w_1,w_2,y_1])^2\mathbb P(w_1,w_2,y_1) dw_1dw_2dy_1\nonumber\\
    &=\sum_{w_2\in\{a,-a\}}\iint (\mathbb E[X_1|w_1,w_2,y_1])^2\mathbb P(w_1,w_2,y_1) dw_1dy_1\nonumber\\
    &=\iint (\mathbb E[X_1|w_1,w_2=a,y_1])^2\mathbb P(w_1,w_2=a,y_1) dw_1dy_1 + \iint (\mathbb E[X_1|w_1,w_2=-a,y_1])^2\mathbb P(w_1,w_2=-a,y_1) dw_1dy_1\nonumber\\
    &=\iint w_1^2 \cdot \sbrackets{\mathbb P(w_1,w_2=a,y_1) + \mathbb P(w_1,w_2=-a,y_1)} dw_1dy_1+ a\iint G_0\sbrackets{(1-\gamma)G_1 - \gamma G_2}w_1dw_1dy_1\nonumber\\
    &\quad + a\iint G_0\sbrackets{\gamma G_1 - (1-\gamma) G_2}w_1dw_1dy_1+ \frac{a^2}{2}\iint \underbrace{G_0\sbrackets{\frac{ [(1-\gamma)G_1 - \gamma G_2]^2}{(1-\gamma)G_1 + \gamma G_2} + \frac{ [\gamma G_1 - (1-\gamma) G_2]^2}{\gamma G_1 + (1-\gamma) G_2}}}_{\triangleq I(w_1,y_1)}dw_1dy_1\nonumber\\
    &= \mathbb E[W_1^2]+a\iint G_0\sbrackets{G_1 -  G_2}w_1dw_1dy_1+ \frac{a^2}{2}\iint I(w_1,y_1)dw_1dy_1.\label{eq: conditional exp}
\end{align}
The first term above boils down to $\mathbb E[W_1^2] = V_1$. Moreover, since
\begin{align*}
   &\iint G_0G_1w_1dw_1dy_1\\
    &=\iint \frac{1}{\sqrt{V_1}} \phi\brackets{\frac{w_1}{\sqrt{V_1}}}\frac{1}{\sqrt{N}} \phi\brackets{\frac{y_1-(w_1+a)}{\sqrt{N}}}w_1dw_1dy_1\\
    &=\int \frac{w_1}{\sqrt{V_1}} \phi\brackets{\frac{w_1}{\sqrt{V_1}}} \sbrackets{\int \frac{1}{\sqrt{N}} \phi\brackets{\frac{y_1-(w_1+a)}{\sqrt{N}}} dy_1}dw_1,
\end{align*}
where the inner integral is with regard to the distribution of $Y_1\sim\mathcal{N}(w_1+a,N)$ conditioned on $W_1=w_1$, therefore the integral value is actually a cumulative distribution with
\begin{align*}
    \int \frac{1}{\sqrt{N}} \phi\brackets{\frac{y_1-(w_1+a)}{\sqrt{N}}} dy_1=1,
\end{align*}
and in this way, the outer integral becomes
\begin{align*}
    \int \frac{w_1}{\sqrt{V_1}} \phi\brackets{\frac{w_1}{\sqrt{V_1}}} dw_1 = \mathbb E[W_1] = 0.
\end{align*}
As a result, we have
\begin{align*}
    \iint G_0G_1w_1dw_1dy_1 = 0.
\end{align*}
Similarly, we can obtain
\begin{align*}
    \iint G_0G_2w_1dw_1dy_1 = 0.
\end{align*}
Therefore, the term in \eqref{eq: conditional exp} becomes
\begin{align*}
   a\iint G_0\sbrackets{G_1 -  G_2}w_1dw_1dy_1 =0.
\end{align*}
Overall, the MMSE of estimating $X_1 = W_1+a\cdot S$ given $W_1,W_2,Y_1$ is 
\begin{align*}
    \mathbb E[(X_1-\mathbb E[X_1|W_1,W_2,Y_1])^2]
    &= \mathbb{E}[X_1^2] - \mathbb{E}\sbrackets{\left(\mathbb{E}\sbrackets{X_1|W_1,W_2,Y_1}\right)^2}\\
    &=\mathbb E[(W_1+a\cdot S)^2] - \mathbb{E}\sbrackets{\left(\mathbb{E}\sbrackets{X_1|W_1,W_2,Y_1}\right)^2}\\
    &=V_1+a^2-\brackets{V_1  +\frac{a^2}{2}\iint I(w_1,y_1)dw_1dy_1}\\
    &=a^2 -\frac{a^2}{2}\iint I(w_1,y_1)dw_1dy_1.
\end{align*}
This concludes the proof of Theorem \ref{thm: coord_2_testch}

\end{proof}

\end{document}